\documentclass{PoS}

\title{Highlights from the VERITAS AGN Observation Program}

\ShortTitle{VERITAS AGN Highlights}

\author{\speaker{Wystan Benbow} for the VERITAS Collaboration\thanks{http://veritas.sao.arizona.edu/}\\
        Harvard-Smithsonian Center for Astrophysics\\
        E-mail: \email{wbenbow@cfa.harvard.edu}\\}

\abstract{The VERITAS array of four 12-m imaging atmospheric-Cherenkov
  telescopes began full-scale operations in 2007, and is one of the
  world's most sensitive detectors of astrophysical VHE (E>100 GeV)
  $\gamma$-rays. Observations of active galactic nuclei (AGN) are a major 
  focus of the VERITAS Collaboration, and more than 60 AGN, primarily
  blazars, are known to emit VHE photons.  Approximately 3400 hours 
  have been devoted to the VERITAS AGN observation program and 
  roughly 160 AGN are already observed with the array, in most cases 
  with the deepest VHE exposure to date. These observations have
  resulted in 34 detections, most of which are accompanied by
  contemporaneous, multi-wavelength observations, enabling a more 
  detailed study of the underlying jet-powered processes.  Recent 
  highlights of the VERITAS AGN observation program, and the 
  collaboration's long-term AGN observation strategy, are presented. }

\FullConference{The 34th International Cosmic Ray Conference,\\
		30 July- 6 August, 2015\\
		The Hague, The Netherlands}

\begin{document}

\section{Introduction}
\vspace{-0.4cm}
AGN emit non-thermal radiation across the entire broadband spectrum
and are among the most powerful particle accelerators in the universe.
They are the most numerous class of identified VHE $\gamma$-ray emitter
and comprise approximately one-third of the VHE sky catalog.  As of
June 2015, sixty-two AGN are identified as VHE sources.  These
all belong to the small fraction of AGN which possess jets powered by accretion onto a 
supermassive black hole (SMBH).  It is believed that the $\gamma$-ray
emission is produced in these jets, in a compact region near the SMBH event horizon.

Most ($\sim$95\%) of the VHE $\gamma$-ray emitting AGN are blazars, 
a class of AGN with jets pointed along
the line-of-sight to the observer.  The VHE blazar population includes
four blazar subclasses:  45 high-frequency-peaked BL\,Lac objects (HBLs),
7 intermediate-frequency-peaked BL Lac objects (IBLs), 
1 low-frequency-peaked BL Lac object (LBL),
and 5 flat-spectrum radio quasars (FSRQs), as well as one
gravitationally-lensed blazar whose sub-classification is uncertain.  Although 
the attenuation of VHE photons on the extragalactic background
light (EBL) leads to an eventual distance horizon, blazars are 
detected at VHE out to a redshift of $z = 0.944$.  The only other AGN
detected at VHE are three nearby ($z<0.02$) FR-I radio galaxies.

Empirically, the photon spectra of the observed VHE emission from AGN are often soft 
($\Gamma_{obs} \sim 3 - 5$), and rarely is any emission observed
above $\sim$1 TeV.  This is in part due to the softening of 
the emitted blazar spectra via EBL effects, and is in part due to the VHE
band often being located above the high-energy peak of the typical
double-humped spectral energy distribution (SED) seen from AGN. 
The VHE flux observed from an AGN is almost always variable, similar to its
behavior at other wavelengths.  Typically variations of a factor of 2-3 are
seen on timescales ranging from days to years.  In other cases,
particularly for non-HBL blazars, these variations are what
temporarily made the object detectable at VHE. The detection of rapid
(minute-scale), large-scale (factor of 100) variations of the VHE flux 
remains relatively rare.  

Understanding VHE AGN and their related science relies
on making precision measurements of their spectra,
their variability patterns, and on contemporaneous multi-wavelength (MWL) 
observations.  Typically these studies enable modeling of the AGN SEDs, as well as searches for
correlations in the flux/spectral changes observed that
may indicate commonalities in the origin of the observed emission.
As third-generation VHE facilities such as VERITAS have been operating
for nearly a decade, many of the major breakthroughs in research
related to VHE AGN are now driven by the observation of major flaring episodes.
Correspondingly, the VERITAS AGN program is now largely focused on the
search for, and the observation of, these flares.

\vspace{-0.4cm}
\section{VERITAS AGN Program}
\vspace{-0.3cm}
VERITAS \cite{VERITAS_spec} is most sensitive between $\sim$85 GeV and $\sim$30 TeV and began routine scientific observations with the full array in 
September 2007. VERITAS can perform spectral reconstruction
above $\sim$100 GeV, and has a sensitivity yielding
a 5 standard deviation ($\sigma$) detection of an object with flux
equal to 1\% Crab Nebula flux (1\% Crab) in $\sim$25 hours.  In Summer 2012, a major
multi-step upgrade of VERITAS was completed, significantly improving its low-energy
response.  VERITAS now has $\sim$40\% lower energy threshold, and
detects AGN $\sim$2 times faster than in 2009.

VERITAS acquires $\sim$980 h of good-weather observations each year
during ``dark time'', and observations of AGN average $\sim$425 h per
year. Historically, these data are split $\sim$90\% to blazars,
primarily BL Lac objects, and $\sim$10\% to radio-galaxies, primarily
M\,87.  Beginning in September 2012, the VERITAS
collaboration developed the capability to observed during periods of
``bright'' moonlight (i.e. $>$30\% illumination).  These
higher-threshold data add another $\sim$30\% to overall good-weather
data yield, of which $\sim$200 h per year are used for AGN studies.
AGN comprise 63\% of the VERITAS source catalog and Table~\ref{blazar_table} shows the 34 AGN
detected by VERITAS.  

Target-of-opportunity (ToO) observations have always been, and
continue to be, a key component ($\sim$30\%) of the VERITAS AGN (radio galaxy and blazar)
program. However, the direction of the overall program has evolved
over the years. In general, the philosophy has shifted from an
emphasis on expanding the source catalog by discovering new VHE AGN,
to exploiting the existing catalog via deep / timely measurements of
the known sources. Initially the blazar program was $\sim$80\% (2007-10) devoted to
discovery efforts and following up on any successes.  This effort is
now  $<$20\% of the AGN dark-time program (2012-15).  Similarly the radio-galaxy program was $\sim$40\%
VHE discovery efforts, and no discovery data were taken the past
three seasons.  The VERITAS AGN program is now
heavily devoted to regular VHE monitoring of the entire Northern VHE catalog,
with a cadence designed to generate deep exposures for some
particularly interesting targets, with intense MWL ToO follow-up planned
for any interesting flaring events. These monitoring observations are
coordinated with partners at lower energy, so that long-term
contemporaneous MWL data sets exist for all Northern VHE AGN.

\begin{table}[t]
\begin{center}
\begin{tabular}{c | c | c | c }
\hline
{\footnotesize AGN} & {\footnotesize $z$} &  {\footnotesize Type} & {\footnotesize log$_{10}(\nu_{\rm synch})$ [Hz]}\\
\hline
\vspace{-0.13cm}
{\footnotesize M\,87} & {\footnotesize 0.004} & {\footnotesize FR\,I} & {\footnotesize $--$}\\
\vspace{-0.13cm}
{\footnotesize NGC\,1275} & {\footnotesize 0.018} & {\footnotesize FR\,I} & {\footnotesize $--$}\\
\vspace{-0.13cm}
{\footnotesize Mrk\,421} & {\footnotesize 0.030} & {\footnotesize HBL} & {\footnotesize 18.5}\\
\vspace{-0.13cm}
{\footnotesize Mrk\,501} & {\footnotesize 0.034} & {\footnotesize HBL} & {\footnotesize 16.8}\\
\vspace{-0.13cm}
{\footnotesize 1ES\,2344+514} & {\footnotesize 0.044} & {\footnotesize HBL} & {\footnotesize 16.4}\\
\vspace{-0.13cm}
{\footnotesize 1ES\,1959+650} & {\footnotesize 0.047} & {\footnotesize  HBL} & {\footnotesize 18.0}\\
\vspace{-0.13cm}
{\footnotesize 1ES\,1727+502} & {\footnotesize 0.055} & {\footnotesize  HBL} & {\footnotesize 17.4}\\
\vspace{-0.13cm}
{\footnotesize BL\,Lac} & {\footnotesize 0.069} & {\footnotesize  IBL} & {\footnotesize 14.3}\\
\vspace{-0.13cm}
{\footnotesize 1ES\,1741+196} & {\footnotesize 0.084} & {\footnotesize  HBL} & {\footnotesize 17.9}\\
\vspace{-0.13cm}
{\footnotesize W\,Comae$^{\dagger}$} & {\footnotesize 0.102} & {\footnotesize IBL} & {\footnotesize 14.8}\\
\vspace{-0.13cm}
{\footnotesize RGB\,J0521.8+2112$^{\dagger}$} & {\footnotesize 0.108} & {\footnotesize HBL} & {\footnotesize $--$}\\
\vspace{-0.13cm}
{\footnotesize RGB\,J0710+591$^{\dagger}$} & {\footnotesize 0.125} & {\footnotesize HBL} & {\footnotesize 21.1}\\
\vspace{-0.13cm}
{\footnotesize H\,1426+428} & {\footnotesize 0.129} & {\footnotesize HBL} & {\footnotesize 18.6}\\
\vspace{-0.13cm}
{\footnotesize S3\,1227+25$^{\dagger}$} & {\footnotesize 0.135} & {\footnotesize IBL} & {\footnotesize 14.8}\\
\vspace{-0.13cm}
{\footnotesize 1ES\,0806+524$^{\dagger}$} & {\footnotesize 0.138} & {\footnotesize HBL} & {\footnotesize 16.6}\\
\vspace{-0.13cm}
{\footnotesize 1ES\,0229+200} & {\footnotesize 0.140} & {\footnotesize HBL} & {\footnotesize 19.5}\\
\vspace{-0.13cm}
{\footnotesize 1ES\,1440+122$^{\dagger}$} & {\footnotesize 0.162} & {\footnotesize IBL} & {\footnotesize 16.5}\\
\vspace{-0.13cm}
{\footnotesize RX\,J0648.7+1516$^{\dagger}$} & {\footnotesize 0.179} & {\footnotesize HBL} & {\footnotesize $--$ }\\
\vspace{-0.13cm}
{\footnotesize 1ES\,1218+304} & {\footnotesize 0.184} & {\footnotesize HBL} & {\footnotesize 19.1 }\\
\vspace{-0.13cm}
{\footnotesize RBS\,0413$^{\dagger}$} & {\footnotesize 0.190} & {\footnotesize HBL} & {\footnotesize 17.0}\\
\vspace{-0.13cm}
{\footnotesize 1ES\,1011+496} & {\footnotesize 0.212} & {\footnotesize HBL} & {\footnotesize 16.7}\\
\vspace{-0.13cm}
{\footnotesize MS\,1221.8+2452} & {\footnotesize 0.218} & {\footnotesize HBL} & {\footnotesize 14.0}\\
\vspace{-0.13cm}
{\footnotesize 1ES\,0414+009} & {\footnotesize 0.287} & {\footnotesize  HBL} & {\footnotesize 20.7}\\
\vspace{-0.13cm}
{\footnotesize 3C\,66A$^{\dagger}$} & {\footnotesize  $0.33 < z <  0.41$} & {\footnotesize IBL} & {\footnotesize 15.6}\\
\vspace{-0.13cm}
{\footnotesize PKS\,1222+216} & {\footnotesize 0.432} & {\footnotesize FSRQ} & {\footnotesize $--$}\\
\vspace{-0.13cm}
{\footnotesize PG\,1553+113} & {\footnotesize $0.43 < z < 0.50$} & {\footnotesize HBL} & {\footnotesize 16.5}\\
\vspace{-0.13cm}
{\footnotesize PKS\,1424+240$^{\dagger}$} & {\footnotesize $z > 0.604$} & {\footnotesize IBL} & {\footnotesize 15.7}\\
{\footnotesize PKS\,1441+25} & {\footnotesize 0.939} & {\footnotesize FSRQ} & {\footnotesize $--$}\\
\hline
\hline
\vspace{-0.13cm}
{\footnotesize 1ES\,0033+595} & {\footnotesize ?} & {\footnotesize HBL} & {\footnotesize 18.9}\\
\vspace{-0.13cm}
{\footnotesize 1ES\,0502+675$^{\dagger}$} & {\footnotesize ?} & {\footnotesize HBL} & {\footnotesize 19.2}\\
\vspace{-0.13cm}
{\footnotesize 1ES\,0647+250} & {\footnotesize ?} & {\footnotesize HBL} & {\footnotesize 18.3}\\
\vspace{-0.13cm}
{\footnotesize B2\,1215+30} & {\footnotesize ?} & {\footnotesize IBL} & {\footnotesize 15.6}\\
\vspace{-0.13cm}
{\footnotesize HESS\,J1943+213} & {\footnotesize ?} & {\footnotesize HBL} & {\footnotesize $--$}\\
{\footnotesize RGB\,J2243+203$^{\dagger}$} & {\footnotesize ?} & {\footnotesize HBL} & {\footnotesize 14.2}\\
\hline
\end{tabular}
\vspace{-0.2cm}
\caption{{\footnotesize The 34 AGN (25 HBL, 7 IBL, 2 radio galaxies)
    detected with VERITAS. This catalog has grown by 7, 13, and 22 AGN since the
    ICRCs in 2013, 2011 and 2009, respectively.  The
    12 blazars discovered at VHE by VERITAS are marked with a dagger. The
    classifications and synchrotron peak
    frequencies are taken from TeVCat and \cite{Nieppola}, respectively.}}
\label{blazar_table}
\end{center}
\vspace{-0.9cm}
\end{table}

\vspace{-0.4cm}
\section{Highlights from Blazar Discovery Observations}
\vspace{-0.3cm}
Although the VERITAS VHE blazar discovery program has ramped down,
there are a few recent highlights from it.  These include the discovery of VHE emission from two blazars
during ToO observations, and the
preparation of limits for a sample of 114 blazars and 2FGL objects.

{\bf RGB\,J2243+203} is a {\it Fermi}-LAT-detected BL Lac object with unknown
redshift.  It has been classified as an IBL and an HBL, and is probably a borderline case.
The favorable extrapolation of its hard LAT spectrum ($\Gamma_{2FGL} \sim1.75$, $\Gamma_{1FHL} \sim
2.4$) motivated $\sim$5 h of observation of this object by VERITAS in
2009, but it was not detected, yielding a flux limit of $\sim$2\% Crab.  Although previous observations were unsuccessful, this
object was one of hundreds of candidate VHE emitters whose {\it Fermi}-LAT
flux and photon index is automatically monitored on timescales ranging from 0.25 -
28 days by VERITAS.  On December 20, 2014, this pipeline issued a
flaring alert based on a relative flux change (important for HBLs) which initiated VERITAS ToO observations.  It is notable
that the triggering flux was below the threshold typically used by the
{\it Fermi}-LAT team for alerts.  The
VERITAS observations yielded $\sim$4 h of good-quality data over the
following four nights, which were effectively the end of the observing
season for the target.  An excess of $\sim$200 $\gamma$-rays
($\sim$6$\sigma$) was detected from RGB\,J2243+203.  
The observed flux above 160 GeV is $\sim$6\%
Crab and the observed photon index is $\Gamma = 4.6 \pm
0.6$.  More details on these studies are presented in \cite{Udara_ICRC2015}.

\begin{figure*}[!t]
   \centerline{ {\includegraphics[width=2.0in]{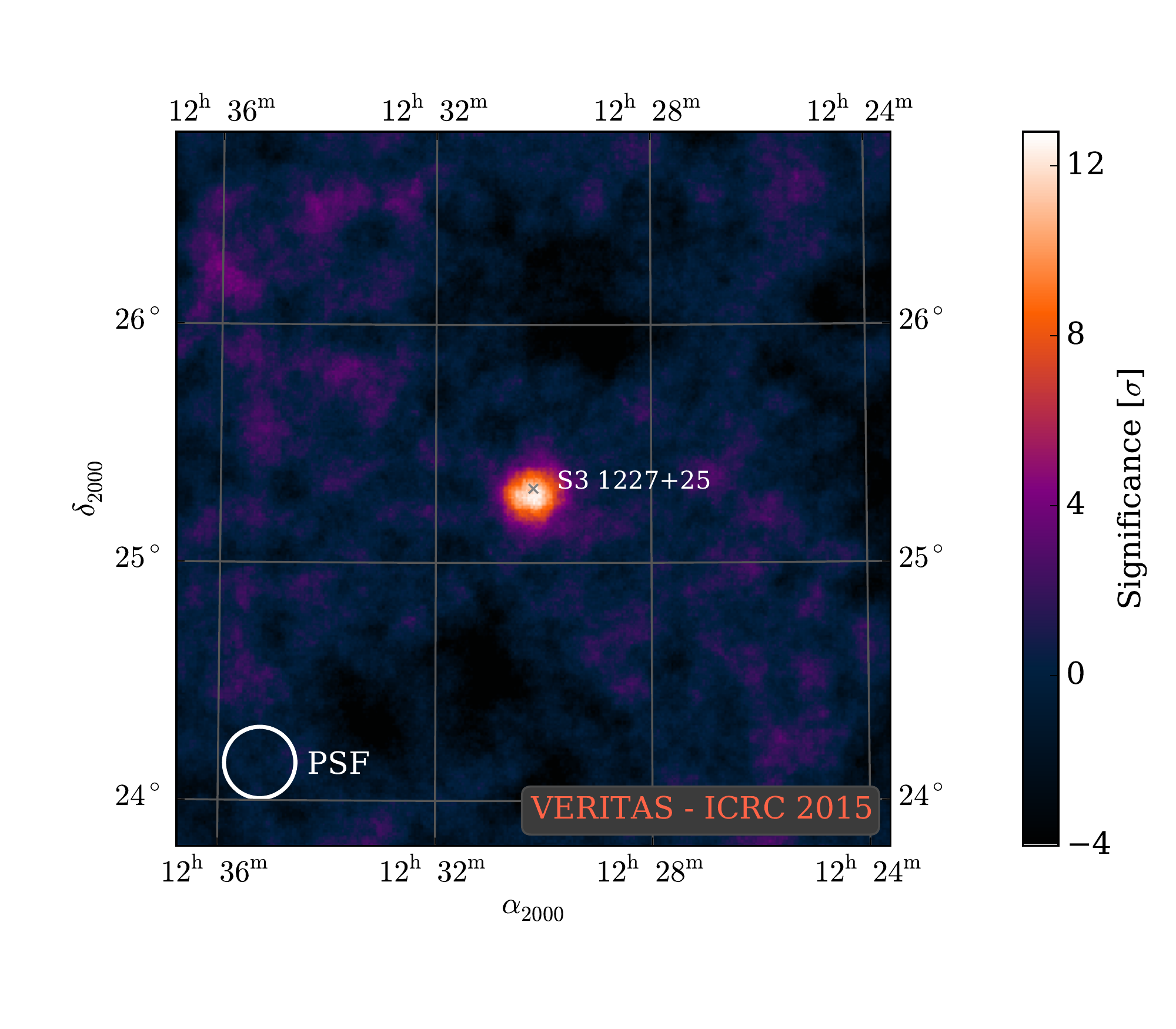} }
              \hfil
              {\includegraphics[width=2.0in]{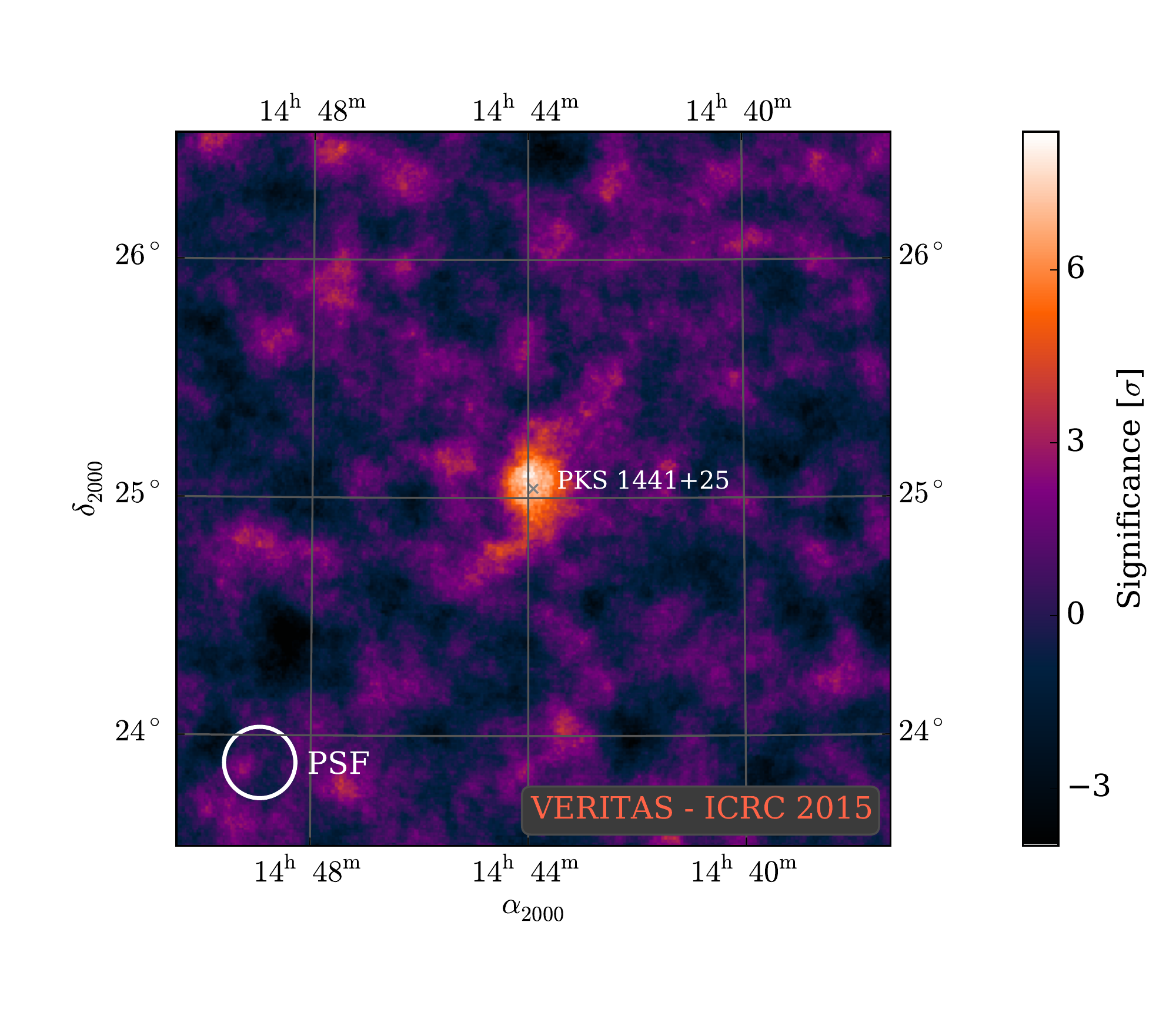} }
              \hfil
              {\includegraphics[width=1.7in]{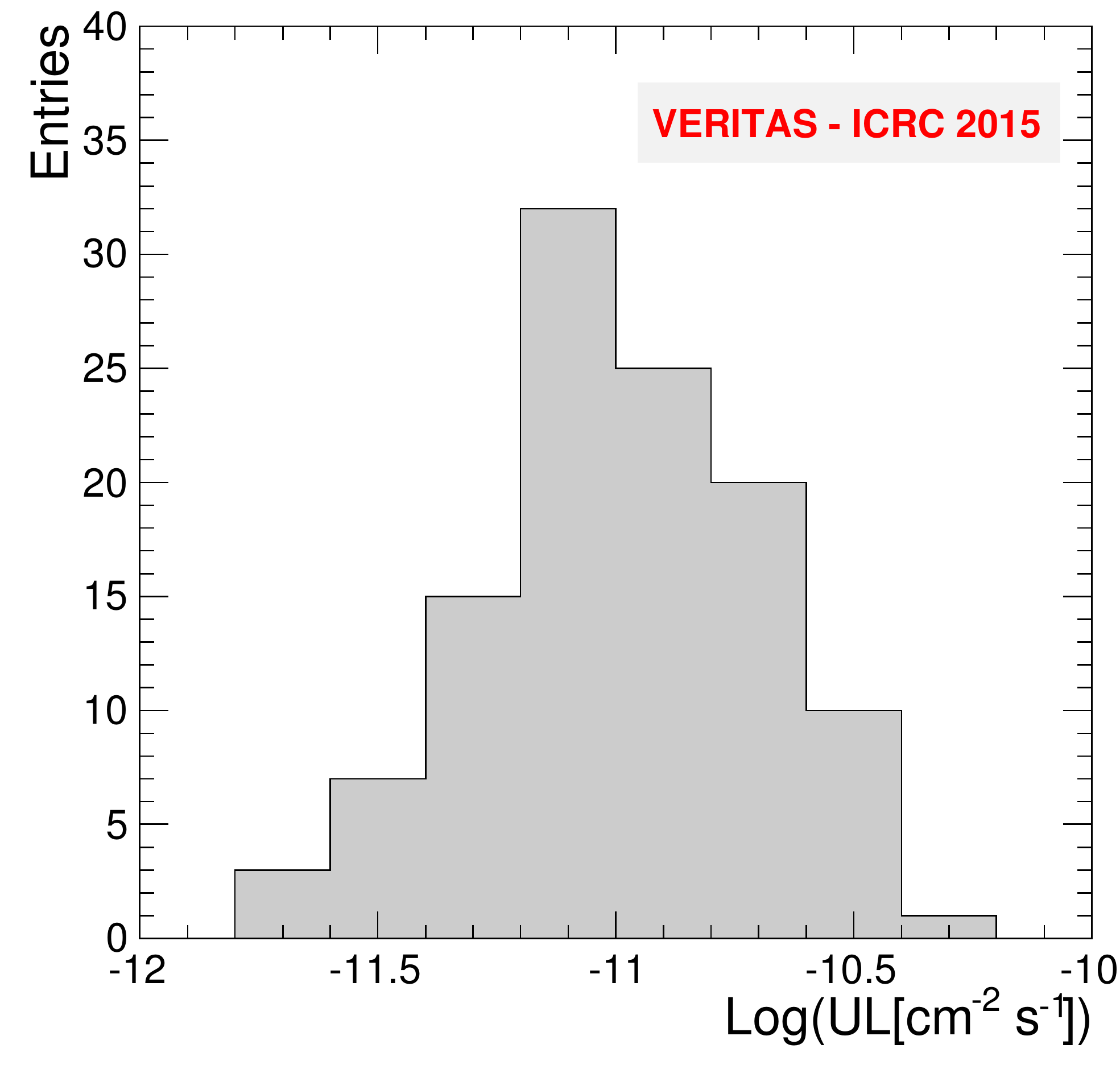} }
             }
   \caption{{\footnotesize The preliminary sky maps of the significance measured from
     the direction of S3\,1227+25 (left) and PKS 1441+25
     (center). Right) The preliminary histogram of integral flux upper limits derived from
     VERITAS observations of 93 blazars and all 21 other 2FGL sources in
     the field of view of those blazars.}}
   \label{Blazar_Maps}
\vspace{-0.2cm}
 \end{figure*}

{\bf S3\,1227+25} (ON\,246) is a {\it Fermi}-LAT-detected IBL at a redshift of
$z = 0.135$.
Its MeV-GeV properties ($\Gamma_{3FGL} \sim2.2$, $\Gamma_{1FHL} \sim
3.3$) suggest it is an unlikely VHE emitter during its baseline emission
state.  In January 2015, the object flared initiating alerts from both
the LAT team and the automated VERITAS pipeline.  In response, VERITAS acquired
$\sim$3 h of good-weather ToO observations, but did not detect the
source despite the reports of $\sim$30 times brighter MeV-GeV flux ($\Gamma_{LAT} = 2.2$).  In May 2015, an even brighter
flare ($\sim$40x flux, $\Gamma_{LAT} = 1.9$) of this blazar initiated similar alerts,
triggering another VERITAS campaign lasting about one week.  Overall, an excess of 420 $\gamma$-rays ($\sim$13$\sigma$)
was detected from S3\,1227+25.  Figure~\ref{Blazar_Maps} shows the sky
map of the significance for the region surrounding S3\,1227+25.  The blazar is clearly detected on three
separate nights: May 16, 18, and 21, 2015 (UTC), noting that weather
issues affected the temporal coverage.  A
variable, soft-spectrum flux was observed, with peaks of  6-8\% Crab above 100 GeV.

{\bf Upper Limits} were derived from VERITAS observations (2007-2012) of 93
blazars that were not eventually detected as VHE emitters.  
The observations of these objects comprise a total good-quality live time of $\sim$570 hours.
While none of these individual sources were detected, a 4.6$\sigma$ (pre-trials)
excess is seen if one stacks the observations of all 36 relatively nearby ($z<0.6$) HBLs in the sample, i.e. the dominant population of extra-galactic
VHE sources.  No significant excess is seen (0.6$\sigma$) by stacking the remaining targets.  
Figure~\ref{Blazar_Maps} shows the distribution of integral flux upper
limits derived from VERITAS observations of each of these blazars, as
well as for the 21 2FGL sources serendipitously located in the
3.5$^{\circ}$ field of view of VERITAS. The typical limit from this
sample is $\sim$2\% Crab, and often the most sensitive produced at VHE.

\vspace{-0.4cm}
\section{Highlights from Observations of Known VHE Blazars}
\vspace{-0.3cm}
Observations of known VHE blazars are the dominant component of the
VERITAS AGN program.  These data include various long-term flux monitoring
programs, ToO observations and single-season campaigns.  A few recent highlights are given here,
as well as some results that have further guided the development of the VERITAS 
blazar observing program.

{\bf PKS\,1441+25} is an FSRQ located at a redshift of $z = 0.939$.
Immediately after the MAGIC discovery (ATel \#7402) of VHE $\gamma$-ray emission from this
object in April 2015, VERITAS initiated a ToO observation campaign.  A
total of 15 h of good-quality observations was acquired during a 1-week
period, resulting in the detection of a very soft spectrum excess of $\sim$400 events
($\sim$8$\sigma$; see Figure ~\ref{Blazar_Maps}).  The observed flux was steady at $\sim$5\% Crab
above 80 GeV.  A further $\sim$4 h of data were taken in May 2015,
after the MeV-GeV ({\it Fermi}-LAT) flare that initiated the MAGIC observations had
subsided.  No significant excess was observed from the source,
unlike the VERITAS detection ($\sim$6$\sigma$ in $\sim$6 h over 10
nights) of steady emission ($\sim$3\% Crab) from the FSRQ PKS 1222+216
approximately two weeks after a similar MeV-GeV flare had subsided in
February 2014.

{\bf 1ES\,0229+200} was observed by VERITAS for 54.3 h from 2009-2012
as part of a MWL monitoring campaign including {\it Swift}, RXTE and
{\it Fermi}-LAT \cite{1ES0229_paper}.  The blazar
was solidly detected by VERITAS ($\sim$12$\sigma$) and the
VHE spectrum well measured ($\Gamma = 2.59 \pm 0.12$).  A variable VHE flux was observed presenting
challenges to efforts to exploit the VHE spectrum of 1ES\,0229+200 to
constrain the strength of the intergalactic magnetic field.  In addition, a synchrotron
self-Compton (SSC) model was fit to the SED.  Rather than presenting a single, degenerate solution the entire range of
SSC parameters allowed by the data were presented.  Although no
detection was particularly strong in any given waveband, in many cases the SSC parameters were
constrained to a factor of $\sim$2 showing the general promise for
VERITAS MWL campaigns on other blazars.

{\bf PKS\,1424+240} is an IBL discovered at VHE by VERITAS in 2009 \cite{PKS1424_paper1}.
In 2013, archival HST measurements were used to show it has $z>0.604$, making it
one of the most distant VHE blazars \cite{PKS1424_HST}.  This
motivated a deep campaign on this object in 2013 \cite{PKS1424_paper2}.
In total more than 100 h of observations, including
archival data, resulted in a $\sim$18$\sigma$ detection above 120 GeV
and a VHE spectrum that is now well measured ($\Gamma = 4.2 \pm 0.3$).  When EBL
effects are removed assuming the redshift is at the lower limit, the spectrum shows an
indication of spectral hardening that is curious and becomes more so with
higher redshift assumptions.  Unfortunately the VHE flux
was more than 2 times lower in 2013 than in 2009, and has remained low since, thus an improvement in the high-energy statistics further
probing this unusual spectral signature has remained difficult.

{\bf HESS\,J1943+213} is a point-like, $>$500 GeV emitter found in the
HESS Galactic Plane scan \cite{HESSJ1943}.    It is a hard-spectrum
{\it Fermi}-LAT source and might be a blazar, but no flux variations have been observed, nor has any
redshift been measured.  To probe the object's possible blazar origins, VERITAS observed HESS\,J1943+213 for 22.5 h
of good-quality live time in 2014.  A strong excess (18$\sigma$;  3.7$\sigma$ h$^{-0.5}$) was
observed, corresponding to a flux that is consistent with the HESS (2005-08)
measurement.  No flux or spectral variations are seen on daily or
weekly time scales.  The observed spectrum ($\Gamma = 2.82\pm0.13$)
is harder than, albeit consistent with, the HESS spectrum ($\Gamma = 3.1\pm0.3)$, whose softness was a key motivator for the blazar
interpretation of this VHE source.  More details on these VERITAS
studies can be found in \cite{1943_ICRC2015}.

{\bf 1ES\,1727+502} was detected by VERITAS for the first time during a flare in May
2013 \cite{1727_paper}.  The observed peak integral flux ($\sim$10\% Crab) is about five times higher 
than the archival VHE flux.  The detection is
notable since it was achieved using observations with a
reduced-high-voltage configuration that enables observations under
bright moonlight.  This detection led to the targets of the VERITAS bright moonlight program being
expanded from a few, generally bright, hard-VHE-spectrum blazars to a
wide variety of targets, including VHE discovery candidates.

{\bf A Snapshot Program} was developed by the VERITAS
collaboration on a trial basis during the 2013-14 season.  Here, a 15-minute observation was taken
each month during the four best months of visibility for each of $\sim$20
VHE blazars not otherwise studied or monitored by VERITAS.  This snapshot
enables VERITAS to detect an excess from a $>$10\% Crab source, which in
most cases would correspond to a bright VHE flare from the object.  While this program
would likely miss any short-duration flaring events (e.g. $<$1-day
flares), the hope was that it would likely catch many longer-duration events
that could be followed up with coordinated MWL campaigns.
This program proved immensely successful.  During the 2013-14 season,
five flaring blazars were detected
by the snapshot program, including 1ES\,0033+595 ($\sim$20$\sigma$, $\sim$15\%
Crab), 1ES\,1011+496 (peaking at $\sim$70\% Crab), MS\,1221.8+2452
($\sim$15\% Crab), VER\,J0521+211 (peaking at $\sim$40\% Crab; see \cite{VERJ0521_ICRC2015}), and 1ES
1727+502 ($\sim$10\% Crab).  These flares complemented
other flares found via VERITAS' intense blazar monitoring
programs  (e.g. a 1-day VHE $\sim$15\% Crab flare of BL Lac in 2013; a
$>$200\% Crab flare of B2\,1215+30 -- see \cite{1215_ICRC2015}), or via ToO
observations triggered by events at lower energy (e.g. PKS\,1222+216
in 2014).  Following up on this trial, a more intense
{\it Snapshot Program} was carried out in 2014-15 (see below).  Although the results
were less spectacular, elevated states were found from several of the
targets.

\vspace{-0.4cm}
\section{Long-term Blazar Observing Strategy}
\vspace{-0.3cm}
In Fall 2014, a new long-term (5-year) strategy for the VERITAS AGN observation
program during dark-time was implemented.  While the program will
naturally evolve, the following describes the current strategy.  For blazars, deep observations
via intense, regular monitoring will be taken for 10 blazars fitting into three
core programs (128 h total).  One core program will focus on {\it EBL and Cosmic-ray
Line of Sight Measurements} via observations of five moderately
distant, hard VHE spectrum
blazars (1ES\,0229+200, RGB\,J0710+591, 1ES\,1218+304,
PKS\,1424+240 and H\,1426+428).  Another core program will focus on
{\it Understanding MWL Variations} via observations of three
blazars that are highly variable at all wavelengths (3C\,66A, W\,Comae and
BL\,Lac),  Another core program will focus on {\it Iconic Objects} by
generating regular single-night spectra for Mkn\, 421 and Mkn
421, as components of a major MWL effort.
Another key component of the long-term strategy is the {\it Snapshot Program}.
Here,  VERITAS will take weekly snapshot observations of any other visible
targets remaining in the Northern VHE blazar catalog (36 targets), to detect flaring events (program
total = 86 h).  While the duration of each target's snapshot varies
(typically 15 min, up to 1 h), the minimal detection sensitivity of any snapshot is
10\% Crab flux.  An automatic, real-time analysis pipeline and
a comprehensive decision tree for ToO-triggering ensures
that instantaneous follow-up of any flare at least five times the
base-line flux (minimally 10\% Crab) occurs.  Each year an 80-h allocation
is pre-approved for ToO follow-up of flaring events meeting any of a number of MWL /
VHE triggers; the pre-approval exists to reduce logistical hurdles for triggering, including instantaneous ones, 
and additional ToO time is possible.  The aforementioned monitoring
observations are taken simultaneously with {\it Swift} UVOT / XRT for the 10
core-program blazars, as well as for 7 other high-priority blazars.  For each of the 46 blazars intense
BVri coverage is arranged with the FLWO 48-inch optical telescope.
Additional time ($\sim$100 h / year) is envisioned for {\it Other Blazar
Projects}.

\begin{figure}[t]
  \vspace{0.0cm}
\centering
\includegraphics[width=4.in]{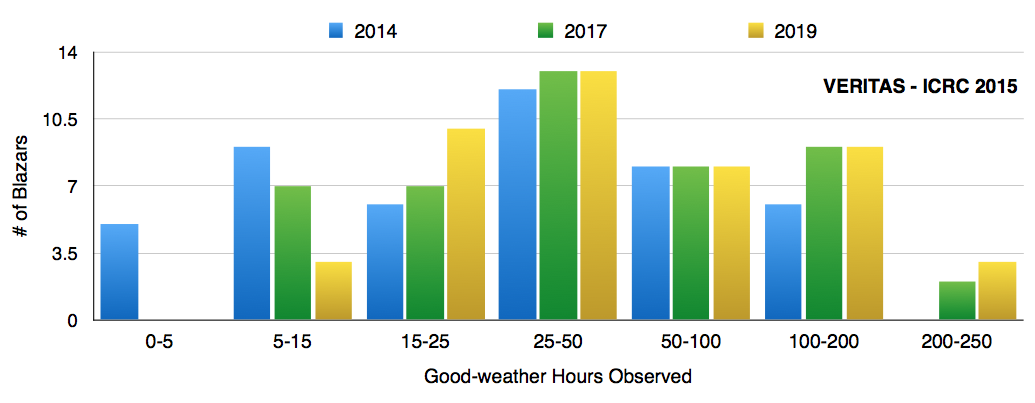}
\vspace{-0.2cm}
\caption{{\footnotesize The histogram of the actual, or minimally expected, VERITAS exposure
    for each Northern Hemisphere blazar in July of 2014 (mean = 32 h;
    median = 46 h), 2017 (mean = 39 h; median = 61 h) and
    2019 (mean = 44 h; median = 71 h).  No observations beyond the monitoring programs (e.g. ToO
    data) are assumed.  Three blazars not known to be VHE
    emitters in July 2014 are included.}}
  \label{Min_exposure}
\vspace{-0.3cm}
 \end{figure}

Overall, this program should provide regular sampling of the light
curves of all Northern-Hemisphere VHE blazars for a five year period,
and in 15 cases for a $\sim$10-year period, all with intense MWL coverage.  In addition, every VHE
blazar visible to VERITAS will eventually have reasonably deep coverage by 2019.
Figure~\ref{Min_exposure} shows a histogram of the minimum total
VERITAS exposure for each of the Northern-Hemisphere VHE blazars
that will be acquired under this plan by July 2017 and July 2019, as
well as where the exposures stood in July 2014.  

\vspace{-0.4cm}
\section{Recent Radio Galaxy Studies \& Future Plans}
\vspace{-0.3cm}
The VERITAS radio-galaxy program has long consisted of deep monitoring
of the known VHE emitter M\,87, ToO observations on the known VHE
emitter NGC\,1275 (since 2009),  and discovery observations of a few EGRET- and
{\it Fermi}-LAT-detected objects.  Recently the discovery component was converted
to a {\it Fermi}-LAT-based ToO program.  The M\,87 program has been very
successful, with $\sim$230 h of good-weather data being acquired, along
with intense MWL coverage and the observation of bright flares in 2008 \cite{M87_flare1}
and 2010 \cite{M87_flare2}.  Recently the AGN has been fairly quiet aside from a
relatively minor brightening in 2012 \cite{Beilicke_HDGS12}.  In the past two seasons, the
target was only weakly detected ($\sim$6$\sigma$) in modest exposures
($<$16 h good-quality live time) taken each season, with the observed flux near the baseline value ($\sim$1.5\% Crab).
NGC 1275 was recently detected ($\sim$7$\sigma$) by VERITAS during ToO observations
taken in December 2012 - February 2013 ($\sim$16 h) and in  October -
November 2013  ($\sim$16 h).  A soft-spectrum ($\Gamma \sim 4$; 1\%
Crab) was observed, consistent with prior MAGIC measurements.
In the future, M\,87 will continue to be regularly monitored (20 h / yr) as
part of a major world-wide effort, and
ToO observations will be taken of NGC\,1275, as well as any {\it Fermi}-LAT-detected
radio galaxy should the LAT flux exceed a prescribed threshold. 

\vspace{-0.4cm}
\section{Conclusions}
\vspace{-0.3cm}
AGN observations remain a major component ($\sim$50\%) of the 
scientific program of VERITAS.  Thirty-two BL Lac objects, two FSRQs
and two FR-I radio galaxies are detected with the observatory, and
constraining upper limits exist from VERITAS observations of
$\sim$100 other AGN.  A strategy guiding
the VERITAS AGN program through 2019 is organized.
This strategy is heavily focused on regular VHE and MWL 
monitoring of all known VHE AGN in the Northern Hemisphere, and
intense ToO follow-up of interesting flaring events.
Although the VERITAS AGN discovery program has decreased in scope,
the collaboration continues this effort, largely with ToO observations
during ``dark time'' and non-ToO bright-moon observations.  Given the major
commitment to AGN research, there should be
many exciting VERITAS results still to come.

\vspace{0.1cm}
{\footnotesize
This research is supported by grants from the U.S. Department of
Energy Office of Science, the U.S. National Science Foundation and the
Smithsonian Institution, and by NSERC in Canada. We acknowledge the
excellent work of the technical support staff at the Fred Lawrence
Whipple Observatory and at the collaborating institutions in the
construction and operation of the instrument.  The VERITAS
Collaboration is grateful to Trevor Weekes for his seminal
contributions and leadership in the field of VHE gamma-ray
astrophysics, which made this study possible.}

\end{document}